\input harvmac
%\draftmode

\def \om {\omega}

\def \k {\kappa} 

\def \g {\gamma}
\def \del {\partial}

\def \ha{{\textstyle{1\over 2}}}

\def \a {\alpha}

\def \chi {\chi}

\def \p {\phi}
\def \m {\mu}

\def \l {\lambda}

\def \td {\tilde }
\def \d {\delta}

\def \inv {^{-1}}
\def \ov {\over }

\def \lr { \lref}
\def\np {{  Nucl. Phys. }}
\def \pl {{  Phys. Lett. }}
\def \mpl {{ Mod. Phys. Lett. }}
\def \prl {{  Phys. Rev. Lett. }}
\def \pr  {{ Phys. Rev. }}
\def \ap  {{ Ann. Phys. }}

\def \cqg {{ Class. Quant. Grav. }}

\baselineskip8pt
\Title{
\vbox
{\baselineskip 6pt{\hbox{PUPT-1613}}{\hbox
{Imperial/TP/95-96/39 }}{\hbox{hep-th/9604089}} {\hbox{
  }}} }
{\vbox{\centerline {Entropy of Near-Extremal Black p-branes  }
% \centerline {           }
\vskip4pt
}}

\vskip -20 true pt

\centerline  { {  I.R. Klebanov\footnote {$^*$} {e-mail address:
 klebanov@puhep1.princeton.edu} }}

 \smallskip \smallskip

\centerline{\it Joseph Henry Laboratories }
\smallskip

\centerline{\it   Princeton Unversity, Princeton, NJ 08544 }

\medskip
\centerline {and}
\medskip
\centerline{   A.A. Tseytlin\footnote{$^{\star}$}{\baselineskip8pt
e-mail address: tseytlin@ic.ac.uk. \ 
On leave  from Lebedev  Physics
Institute, Moscow.} }

\smallskip\smallskip
\centerline {\it  Theoretical Physics Group, Blackett Laboratory,}
\smallskip

\centerline {\it  Imperial College,  London SW7 2BZ, U.K. }
\bigskip\bigskip
\centerline {\bf Abstract}
\medskip
\baselineskip8pt
\noindent
We carry out a thorough survey of entropy for a large class 
of $p$-branes in various dimensions. We find that the Bekenstein-Hawking
entropy may be given a simple world volume interpretation only for the
non-dilatonic $p$-branes, those with the dilaton constant throughout
spacetime. The entropy of extremal 
non-dilatonic $p$-branes is non-vanishing only
for solutions preserving 1/8 of the original supersymmetries.
Upon toroidal compactification these reduce to dyonic black holes
in 4 and 5 dimensions.
For the self-dual string in 6 dimensions, which preserves 1/4 of
the original supersymmetries, the near-extremal entropy is found to agree
with a world sheet calculation, in support of the existing literature.
The remaining 3 interesting cases preserve 1/2 of 
the original supersymmetries. These are the self-dual 3-brane in 10 
dimensions, and the 2- and 5-branes in 11 dimensions.
For all of them the scaling of the near-extremal Bekenstein-Hawking
entropy with the Hawking temperature is in agreement with a statistical 
description in terms of free massless fields on the world volume.

\medskip
%%%%%%%%%%%%%%%%%%%%%%%%%%%%%%%%%%%%%%%%%%%%%%%%%%%%%%%%%
\Date {April 1996}
%%%%%%%%%%%%%%%%%%%%%%%%%%%%%%%%%%%%%%%%%%%%%%%%%%%%%%%%%%%%%%%%%%%
\noblackbox
\baselineskip 14pt plus 2pt minus 2pt
%\baselineskip 20pt plus 2pt minus 2pt
%%%%%%%%%%%%%%%%%%%%%%%%%%%%%%%%%%%%%%%%%%%
\lr\witten{E. Witten, hep-th/9510135.}
\lr\wit{E. Witten, Contribution to Strings '95, hep-th/9507121.}
\lr \row {R. Rohm and E. Witten, \ap 170 (1986) 454.}
\lr\dai{J.~Dai, R.G.~Leigh, and J.~Polchinski,
Mod. Phys. Lett.  A4 (1989) 2073; R.G.~Leigh,
Mod. Phys. Lett. A4 (1989) 2767.}
\lr\polch{
J.~Polchinski, Phys. Rev. Lett. 75 (1995)
4724,  hepth/9510017.}
\lr \dgh {A. Dabholkar, G.W. Gibbons, J. Harvey and F. Ruiz Ruiz,  \np
B340 (1990) 33;
A. Dabholkar and  J. Harvey,  \prl
63 (1989) 478.
}
\lr\mon{J.P. Gauntlett, J.A. Harvey and J.T. Liu, \np B409 (1993) 363.}
\lr\chs{C.G. Callan, J.A. Harvey and A. Strominger, 
\np { B359 } (1991)  611; in {\it 
Proceedings of the 1991 Trieste Spring School on String Theory and
Quantum Gravity}, J.A. Harvey {\it et al.,}  eds. (World Scientific, 
Singapore
1992).}
\lr\hull{C. Hull, hep-th/9512181.}
\lr\vafa{C. Vafa, hep-th/9602022.}
\lr\CM{ C.G. Callan and  J.M.  Maldacena, 
PUPT-1591,  hep-th/9602043.} 
\lr\SV {A. Strominger and C. Vafa, HUTP-96-A002,  hep-th/9601029.}

\lr\MV {J.C. Breckenridge, R.C. Myers, 
A.W. Peet  and C. Vafa, HUTP-96-A005,  hep-th/9602065.}
\lr\breck{J.C. Breckenridge, D.A. Lowe, R.C. Myers,
A.W. Peet, A. Strominger and C. Vafa, hep-th/9603078.}
\lr\kut{D. Kutasov and E. Martinec, hep-th/9602049.}
\lr\mlst{G.T. Horowitz, J.M.  Maldacena and A. Strominger, hep-th/9603109.}
\lr \CT{M. Cveti\v c and  A.A.  Tseytlin, 
\pl { B366} (1996) 95, hep-th/9510097. 
}
\lr \CTT{M. Cveti\v c and  A.A.  Tseytlin, 
IASSNS-HEP-95-102, hep-th/9512031. 
}
\lr\LW{ F. Larsen  and F. Wilczek, 
PUPT-1576,  hep-th/9511064.    }
\lr\TT{A.A. Tseytlin, \mpl A11 (1996) 689, hep-th/9601177.}
\lr \HT{ G.T. Horowitz and A.A. Tseytlin,  \pr { D51} (1995) 
2896, hep-th/9409021.}
\lr\khu{R. Khuri, \np B387 (1992) 315; \pl B294 (1992) 325.}
\lr\CY{M. Cveti\v c and D. Youm,
 UPR-0672-T, hep-th/9507090; UPR-0675-T, hep-th/9508058; 
  \pl { B359} (1995) 87, 
hep-th/9507160.}

\lr\aat{A.A. Tseytlin, \np B469 (1996) 12, hep-th/9602064.}
\lr\ght{G.W. Gibbons, G.T. Horowitz and P.K. Townsend, \cqg 12 (1995) 297,
hep-th/9410073.}
\lr\dul{M.J. Duff and J.X. Lu, \np B416 (1994) 301, hep-th/9306052. }
\lr\hst {G.T. Horowitz and A. Strominger, hep-th/9602051.}
\lr\dull{M.J. Duff and J.X. Lu, \pl B273 (1991) 409. }
\lr\guv{R. G\"uven, \pl B276 (1992) 49. }
\lr\gubs{S.S. Gubser, I.R.   Klebanov  and A.W. Peet, 
hep-th/9602135.}
\lr\astrom{A. Strominger, unpublished notes.}
\lr \std { M.J. Duff and  K.S. Stelle, \pl B253 (1991) 113.}
\lr \das {S. Das  and S. Mathur, hep-th/9601152.}

\lr\hos{G.T.~Horowitz and A.~Strominger, Nucl. Phys. { B360}
(1991) 197.}
\lr\teit{R. Nepomechie, \pr D31 (1985) 1921; C. Teitelboim, \pl B167 (1986) 69.}
\lr \duf { M.J. Duff, P.S. Howe, T. Inami and K.S. Stelle, 
\pl B191 (1987) 70. }
\lr\duh {A. Dabholkar and J.A. Harvey, \prl { 63} (1989) 478;
 A. Dabholkar, G.W.   Gibbons, J.A.   Harvey  and F. Ruiz-Ruiz,
\np { B340} (1990) 33. }
\lr\mina{M.J. Duff, J.T. Liu and R. Minasian, 
\np B452 (1995) 261, hep-th/9506126.}
\lr\dvv{R. Dijkgraaf, E. Verlinde and H. Verlinde, hep-th/9603126;
hep-th/9604055. }
\lr\ast{A. Strominger, hep-th/9512095.}
\lr\gibb{G.W. Gibbons and P.K. Townsend, \prl  71
(1993) 3754, hep-th/9307049.}
\lr\town{P.K. Townsend, hep-th/9512062.}
\lr\kap{D. Kaplan and J. Michelson, \pr  D53 (1996) 3474, hep-th/9510053.}
\

\lr \dufe{ M.J. Duff, S. Ferrara, R.R. Khuri and J. Rahmfeld,
\pl  B356 (1995) 479, hep-th/9506057.}

\lr \dup{M.J. Duff, H. L\" u  and C.N. Pope, hep-th/9603037.}

\lr\dlp {M.J. Duff, H. L\" u and C.N. Pope, hep-th/9604052.}

\lr\stp{H. L\" u, C.N. Pope, E. Sezgin and K.S. Stelle, \np B276 (1995)  669, hep-th/9508042.}

\lr\lup {H. L\" u and C.N. Pope, hep-th/9512012.}
\lr \duff { M.J. Duff and J.X. Lu, \np B354 (1991) 141. } 

\lr\chs{C.G. Callan, J.A. Harvey and A. Strominger, 
\np { B359 } (1991)  611.}
 \lr \lu{ J.X.  Lu, \pl B313 (1993) 29.}
\lr \CY{M. Cveti\v c and D. Youm,
 \pr D53 (1996) 584, hep-th/9507090.  }
 \lr\kall{R. Kallosh, A. Linde, T. Ort\' in, A. Peet and A. van Proeyen, \pr { D}46 (1992) 5278.} 
\lr\TT{A.A. Tseytlin,  \mpl A11 (1996) 689, hep-th/9601177.}
\lr\CYY{M. Cveti\v c and D. Youm, hep-th/9603100.}
\lr \CT{M. Cveti\v c and  A.A.  Tseytlin, 
\pl { B366} (1996) 95, hep-th/9510097. 
}

\def \sh { {\rm sinh}}
\def \ch { {\rm cosh}}

\def \d {d'}

%%%%%%%%%%%%%%%%%%%%%%%
\newsec{Introduction}
%%%%%%%%%%%%%%%%%%%%%%%%

During the past few months remarkable progress towards a microscopic
understanding of the black hole entropy has taken place. Strominger
and Vafa \SV\ considered a class of 5-dimensional black holes with
non-vanishing Bekenstein-Hawking entropy,
\eqn\bek{S_{BH}={A\over 4 G_N}  \ , }
where $A$ is the volume of the horizon. The entropy of 
Dirichlet brane states \refs{\dai,\polch}
carrying the same set of charges was found to
agree with $S_{BH}$ in the limit appropriate for describing
macroscopic black holes \SV. This result has been generalized in a number
of ways \refs{\CM,\MV}.

The direction that is of immediate relevance to
our paper is the extension of the D-brane entropy counting to near-extremal
black $p$-branes. A system that has been thoroughly investigated
in this regard is the dyonic (self-dual)
black string in $6$ dimensions \refs{\CM,\hst}.\foot{The extremal 
limit of the dyonic string solution was found in \dufe.  Upon compactification
on a circle, its `boosted' equal charge version
 reduces to the black hole considered in  \SV.} 
To leading order away from extremality, the Bekenstein-Hawking entropy
of the dyonic string agrees with the counting of low-energy D-brane 
excitations \refs{\CM,\hst,\breck}. Another interesting laboratory
for near-extremal entropy is the self-dual 
3-brane in 10 dimensions \refs{\gubs,\astrom}.
The feature that the self-dual string in 6 dimensions and the self-dual
3-brane in 10 dimensions share is that these solutions are non-dilatonic:
the dilaton field is constant throughout space-time. A divergence
of the dilaton field at the horizon, that is common for other $p$-branes,
is indeed dangerous because it may lead to a strong coupling problem.
For the 3-brane system
one finds that the Bekenstein-Hawking and the D-brane definitions
of entropy agree on the scaling exponents 
with respect to mass and charge \refs{\gubs,\astrom}.
The Bekenstein-Hawking entropy 
of $n$ coincident 3-branes is reproduced by the statistical mechanics
of $6 n^2$ free massless fermions and bosons in $3+1$ dimensions.
On the other hand, at weak coupling the world volume theory appears to
have $8 n^2$ massless degrees of freedom. While we do not understand
the resolution of this puzzle, we will offer some guesses.\foot{
$n$ coincident 3-branes are described by ${\cal N}=4$ supersymmetric
$U(n)$ gauge theory at the point where the full
non-abelian symmetry is restored. In a sense, at this point monopoles and 
dyons become massless which may affect the counting of low-energy states.
We thank E. Kiritsis
for useful discussions on this issue.} We believe that this 
3-brane puzzle will,
in the end, find an interesting resolution.
In this paper we plan to make similar free field comparisons for other
D-brane systems, hoping to get some information about their world volume
structure.

The primary purpose of this paper is to study the near-extremal
Bekenstein-Hawking entropy, and its possible statistical interpretation,
for a large class of black $p$-branes in various dimensions.
One of our results is that a simple interpretation
of the entropy in terms of massless fields in $p+1$ dimensions
is possible only for the non-dilatonic $p$-branes. In addition to the
previously studied systems, these include the $2$- and $5$- branes in $D=11$.
We hope, therefore, that our studies will shed some light on 
the world volume structure of the M-theory branes.

While this paper was in preparation, we received preprint \dlp\ which
discuss a general class of black
$p$-brane solutions. This class includes the
non-dilatonic $p$-branes \ght\ whose entropy was independently studied by us
and which is the primary subject of this paper. For completeness, however,
in section 2 we review and interpret
the Bekenstein-Hawking entropy of the more general
solutions of \dlp. We find that {\it there are no dilatonic $p$-branes
whose near-extremal 
entropy may be explained by free massless fields on the world 
volume}. This brings us back to the non-dilatonic $p$-branes which
we study in the rest of the paper. 
The results are especially puzzling for the $D=11$ $p$-branes.
While their Bekenstein-Hawking entropies scale with the Hawking
temperature appropriately for the world volume massless field
description, the number of such fields is, in general, not an integer.
Moreover, for $n$ parallel 5-branes this number grows as $n^3$, while
for $n$ parallel 2-branes -- as $n^{3/2}$. We suspect that these
exponents may be interpreted in terms of enhanced symmetry
of coincident $p$-branes. Coincident 5-branes are expected to give
rise to tensionless strings \ast. 
Perhaps this is the reason why the
number of massless degrees of freedom grows faster than the $n^2$
growth associated with restoration of $U(n)$ symmetry found
for the D-branes \witten.

In section 3 we present our approach to charge quantization which allows us 
to normalize the Bekenstein-Hawking entropy for systems involving
$n$ coincident $p$-branes. In section 4 this is compared with the statistical
entropy of $N_p$ massless bosons and fermions in $p+1$ dimensions.
The numbers $N_p$ necessary for complete agreement are deduced and commented 
on. In section 5, we conclude
the paper with some speculations on $5$-branes in $D=12$.

%%%%%%%%%%%%%%%%%%%%%%%%%%%%%%%%%%%%%%%%%%%%%%%
\newsec{Black p-branes and their entropy}
%%%%%%%%%%%%%%%%%%%%%%%%%%%%%%%%%%%%%%%%%%%%%%%%%%%%%%%%
\subsec{Review of black p-brane solutions}
%%%%%%%%%%%%%%%%%%%%%%%%%%%%%%%%%%%%%%%%%%%%%
Black p-brane solutions
in various dimensions with one  scalar field   may be 
described in a universal way as 
extrema of the following action  (see \refs{\hos,\dul,\ght,\stp,\lup,\dlp})
\eqn\act{S= -{1\ov 2\k^2} \int d^D x \sqrt g [ R - 
\ha (\del \p)^2   - {1\ov 2 (d+1)!} e^{a\p} F^2_{d+1}]\ , 
\ \ \ \ D=p+d+3 \ . }
A  general class  of such black  p-brane solutions 
was  found  recently in  \dlp.
This class generalizes, in particular, 
the dilatonic solutions of
\refs{\hos,\dul}  to arbitrary $d$ and $p$ not constrained by 
$N=1$ (see (2.2) below).
\dlp\ also presents dilatonic ($a\neq 0$) generalizations
of the non-dilatonic solutions found in  \ght.  
The metrics found in \dlp\ are `superpositions'
of extremal solutions, parametrized by harmonic functions $H$, 
and the Schwarzschild solutions 
parametrized by function $f$,\foot{Our notation differs somewhat from
that in \dlp.  The relation 
to the radial coordinate  used in  \refs{\hos,\dul,\ght}
is the following: 
$\hat r^d = r^d + r^d_- = r^d H (r) ,  $ \  $ 
 \Delta_- (\hat r) =  H^{-1}(r) , \ 
\Delta_+ (\hat r) =  H^{-1}(r) f(r),$ 
\ $\Delta_\pm (\hat r) =  1- {r^d_\pm\over \hat r^d }$, 
\   $r^d_+ = \m^d \ch^2\g, \ 
r^d_- = \m^d \sh^2\g$ (with $r_\pm \to r_0$ in the extremal limit).
When departing from extremality, fixing the charge
translates into holding $r_+ r_-$ fixed. }
\eqn\met{
d s^2 =  H^\a (r)
\bigg( H^{-N}(r) \big[ - f(r) dt^2 + dy_1^2+...+dy^2_{p}\big]
  +  f\inv (r) dr^2 + r^2 
d\Omega^2_{d+1}\bigg)   ,  }
\eqn\pop{ H= 1 + {r_-^d \ov r^d} \ , \ \ \  \  
f = 1 - {\m^d \ov r^d} \ ,  \ \ \  \ \ \ \ r_-^d \equiv \m^d \sh^2 \g \ ,  }
\eqn\msst{ \a=  {N(p+1) \ov D-2}\ , 
\ \ \ \ \  { N}  \equiv 4 [  a^2 + {2d(p+1)\ov D-2}]\inv  \ . }
The extremal limit corresponds to 
$\m \to 0$ ($f\to 1$) and $  \g \to \infty$ with
$\m^d \sh{2\g} $ kept fixed. 
We are primarily interested in solutions which
are supersymmetric in the extremal limit. For such solutions, 
$N$ is an integer \refs{\stp,\lup}. $N$ may be  interpreted  as 
a measure of `compositeness' of a  configuration, i.e. the number 
of different field strengths or charges that were
set equal to each other
in reducing the action to the form \act\ \refs{\lup,\dlp}.
 For $N=1,2$ or $3$, the fraction of residual
supersymmetry is $1/2^N$; for $N=4$  it  is   1/8  for solutions
with $d=1$ (i.e., the configurations which look like
black holes upon reduction to  $D=4$)    or 1/16 for  
 solutions  with $d=0$.
% (i.e., the configurations that
%reduce to $D=3$ black holes). 

In addition to the metric, \met, the solutions of \dlp\
involve the field strength and the dilaton. Here it is important
whether the solution is purely `magnetic' or dyonic.
For purely `magnetic' solutions,
\eqn\magn{
  F_{d+1} = \ha \sqrt N  d \m^{d} \sh 2\g \  \epsilon_{d+1} \ , \ \ \  \ 
\ \  e^{2\p} = H^{-aN}    \ . }
Related  dyonic solutions are obtained from the `magnetic'
one by performing a duality transformation on the $d$-form field
which leaves the metric invariant but affects the dilaton. We may have
dyons if $D=2p+4$ (i.e. $d=p+1$). A well-known example of this
is 
the Reissner-Nordstrom black hole in $D=4$
($d=p+1=1, \ N=4$).\foot{When embedded into string theory, 
this is the dyonic black hole 
with 4 equal (two electric and two magnetic) charges. 
Though the charges are equal, one needs at least 
two different (electric and magnetic) 
vector fields  since the $p=0, D=4$ self-duality
condition   does not have real solutions. 
This is different from 
the 
 $p=1, D=6$ and $p=3, D=10$  cases:  though  
formally  they may be obtained
as self-dual limits of dyonic  p-brane solutions, 
they may be interpreted as intrinsic solutions of theories
with self-dual $(p+2)$-field strengths, the
self-dual gravity in $D=6$  and type IIB supergravity in $D=10$.
}
For dyons with equal electric and magnetic charge, 
the dilaton is constant ($a$ may be set to zero), and
\foot{The rescaling of the magnetic charge by $1/\sqrt 2$
is necessary in order to get the same expression 
for the  metric in the presence of extra electric charge, equal
to the magnetic one.}
\eqn\seek{
 F_{p+2} = F_{d+1}= {1\ov 2\sqrt 2}  \sqrt N  (p+1) \m^{d} \sh 2\g 
\  (\epsilon_{p+2} + \epsilon^*_{p+2}) \ . 
}
In the special case of $D=2p+4$ where $p$ is odd,
the equal charge dyonic solutions are real self-dual: these are
the self-dual string \refs{\dul,\ght}  in $D=6$ ($d=p+1=2, \ N=2$) and 
the self-dual 3-brane in $D=10$ \refs{\hos,\dull}
($d=p+1=4, \ N=1$). 
We will also be concerned with the 11-dimensional solutions:
the fundamental 2-brane \std\ ($d=6,\ p+1=3, \ N=1$)
and the solitonic 5-brane  \guv\ ($d=3,\ p+1=6, \ N=1$).
These solutions are also non-dilatonic simply because there is no dilaton
in $D=11$.
As shown in \ght, for all these non-dilatonic solutions there exist
Bogomolnyi bounds relating the mass and charge at extremality.
Well-known examples of dilatonic solutions, which will be used in our 
discussion of the entropy, 
are the solitonic and the R-R charged 5-branes \refs{\duff,\chs,\hos} in $D=10$
($d=2, \ p+1=6, \ N=1$).

We shall assume that the internal dimensions $y_i$ 
of a $p$-brane are compactified on a torus with  equal periods  $L$. 
Let us first consider the purely `magnetic' solutions
(the discussion of the purely `electric' case is identical). 
The charge per unit volume and the ADM mass are 
given by \refs{\lu,\dul,\dlp}
\eqn\mess{q_p  = {\omega_{d+1}\ov 2\sqrt 2  \k } \sqrt N  d \m^{d} \sh 2\g
\equiv {\omega_{d+1}\ov \sqrt 2  \k } \sqrt N  d  r^{d}_0 
 \ , }
\eqn\messs{
 M_p  = {\omega_{d+1}\ov 2\k^2 } L^p  \m^d
\big( d+1 + Nd\ \sh^2\g \big) =  {\sqrt 2 q_p \ov  \k  \sqrt N  d }
{ d+1 + Nd\ \sh^2\g  \ov \sh 2\g }  \ ,  
}
where $\omega_{d+1}= 2\pi^{{d\ov 2}+1} /\Gamma({{d\ov 2}+1})$
is the volume of a unit $(d+1)$-sphere.
For self-dual solutions ($d=p+1$), e.g. 
the string in $D=6$ and the 3-brane in $D=10$,  
the mass is the same while the electric and magnetic charges are 
each equal to $1\ov \sqrt 2$ of $q_p$ in \mess.    
The $(D-2)$-volume  of the horizon, located at $r=\m$,  
is  found to be 
\eqn\area{A_p = \omega_{d+1} \m^{d+1}  L^{p} H^{N\over 2}(\m) 
    =     \omega_{d+1}^{-{1\ov d}} L^p ({ 2\sqrt 2  \k q_p \ov  \sqrt N  d
})^{d+1\ov d}
(\sh 2\g)^{- {d+1\ov d}} (\ch \g)^{N}  \ .  }
This volume, and the corresponding Bekenstein-Hawking
entropy, $S_p = 2\pi A_p/\k^2$, have non-vanishing values
in the extremal limit iff  
\eqn\lal{ \l \equiv {d+1\ov d } -  {N\over 2}= 0 \ . }
For supersymmetric $p$-branes (those with integer $N$)
there are only two solutions of this condition:
$d=1, \ N=4$ (i.e. $p=D-4$)  and $d=2, \ N=3$ (i.e. $p=D-5$). 
These respectively correspond to 
the equal charge dyonic black holes in 
$D=4$ \refs{\kall,\CY,\CT}
and $D=5$ \refs{\SV,\TT,\CYY,\mlst}, or
to composite higher-dimensional 
p-branes which reduce to such black holes
upon wrapping over compact dimensions.
This analysis implies, in particular, that 
there are no finite entropy 
extremal black holes in $D=6$ and higher. One  may 
justify this conclusion also  by a  different  argument.
Let us try to construct an extremal supersymmetric $D>5$
black holes using  the `boosted'  fundamental string solution, 
$$ds^2=  H^{-1} (-dt^2 + dy^2) + (dt-dy)^2  +  dx^2_i\ , 
\quad\qquad  e^{2 \phi} =  H^{-1}\ ,$$
and the
R-R $p$-brane solution ($p \leq 4$), 
$$ds^2=  H^{-1/2} (-dt^2 + dy_n^2)   +  H^{1/2} dx^2_i\ , \quad\qquad 
e^{2 \phi} =  H^{(3-p)/2}\ ,$$ 
 as the building blocks (solitonic 5-brane has only $SO(4)$ isometry). 
To get  finite entropy, one  needs a
BPS  `mixture'  with 
the dilaton and all moduli being finite at the horizon. 
The  metric of a  black hole in $D$ dimensions should take the form
(for all charges, i.e. harmonic functions,  set equal)
$$ds^2_D = - f(x) dt^2  +   H^n r^2 d\Omega_{D-2}\ .$$
Since  $H= 1 + m/r^{D-3}$, 
to  get a finite area of the $r=0$  horizon  it is necessary that  
$n= 2/(D-3)$.
This gives 
$n= 2 $ in $D=4$,  \  $ n=1$ in  $D=5$,
but 
$n=2/3$ in $D=6$. 
While for $ n=1,2$ one can build up  $H^n$ from the $H^{1/2}$ factors 
 in the transverse metric of the  R-R $p$-branes, 
 this is impossible 
for $n=2/3$.
 For $D=7$ we need $n=1/2$ but the moduli  blow up at $r=0$.
Thus, it appears that the extreme dyonic black holes in $D=4$
(with 4 charges) and in $D=5$ (with 3 charges)
 are the only ones which have finite 
Bekenstein-Hawking entropy.\foot{ 
Similar conclusion 
about the non-existence of supersymmetric extremal $D> 5$ black holes
with finite area of the horizon was reached by M. Cveti\v c and D. Youm 
(private communication)
who constructed general 
non-extreme rotating BH solutions in $D\geq 6$.}

%%%%%%%%%%%%%%%%%%%%%%%%%%%%%%%%%%%%%%%%%%%%%%%%%
\subsec{The Bekenstein-Hawking entropy of near-extremal black p-branes}
%%%%%%%%%%%%%%%%%%%%%%%%%%%%%%%%%%%%%%%%%%%%%%%%%%%%%%% 

Given that all the 1/2 and 1/4 supersymmetric extremal p-branes 
have zero entropy, it is of interest to study 
their near-extremal  expressions
for the entropy, generalizing previous discussions 
of the $D=6$ dyonic string \hst\  and  
the $D=10$ 3-brane \gubs.
Our eventual goal is 
to interpret the leading near-extremal corrections
to the black $p$-brane mass and entropy 
as the energy and entropy of some weakly interacting 
effective massless  degrees of freedom 
on the $p$-brane world volume. 
In $D\leq 10$ the degrees of freedom are undoubtedly those of the
open strings 
in the D-brane description of the R-R charged $p$-branes.
In the M-theory they are presumably related to  
closed strings inside a 2-brane \kut\ or a 5-brane \dvv. 
Thus, we expect our comparison of entropy
to be a new source of information about
the M-theory.

Let us expand the $p$-brane mass, \messs, for fixed $q_p$ near
extremality (for large $\g$) 
\eqn\lam{ M_p= M_{p0}(1 + {\delta M_p\ov M_{p0}})  \ ,
\ \ \ \ \ \ M_{p0} =\sqrt N  { q_p \ov \sqrt 2 \k }L^p = 
{\omega_{d+1}\ov  2\k^2 } d N r_0^{d} L^p \ , \ \ \ \ 
}
\eqn\pert{
{\delta M_p\ov M_{p0}} = {4 \l \ov N }   e^{-2\g} \ , }
where $\l$ is defined in \lal.
Assuming $\l \geq 0$,
the area of the  horizon, \area, becomes\foot{ For $\l=0$ the 
expression for the entropy given below
applies to the extremal case.
In this special sitation we ignore the subleading corrections
to the  entropy. } 
\eqn\are{
A_p = 4^\lambda\omega_{d+1}^{-{1\ov d}} L^p ({ \sqrt 2  \k q_p \ov  
\sqrt N  d })^{d+1\ov d}
\ e^{- 2 \l  \g}   }
$$
= \ 
\omega_{d+1}^{-{1\ov d}} L^p ({ \sqrt 2  \k q_p \ov  \sqrt N  d })^{d+1\ov d}
({N\delta M_p\ov \l M_{p0}})^\l  \ . 
$$
Interpreting $\delta M_p$ as the energy  $E$, and using \lam,
we obtain the following 
expression for the entropy, 
\eqn\ent{
S_p = {2\pi A_p\ov \k^2} = {4\pi \omega_{d+1}^{-{1\ov d}} 
} L^{p(1-\l)} d^{- {d+1 \ov d}} \l^{-\l} (\sqrt 2  \k)^{{2\ov d} - {N\ov 2}}
  ({  q_p \ov  \sqrt N})^{N\ov 2}  E^\l  \ . }
Remarkably, the dependence of the entropy on the charge 
looks universal --
each of the $N$  `constituent' 1/2 supersymmetric objects making up
the $p$-brane contributes a factor of $\sqrt {q_p}$.

Employing the  thermodynamical relation $dE= T dS$
we find the corresponding Hawking temperature,
\eqn\tem{ 
T\inv ={4\pi \omega_{d+1}^{-{1\ov d}} }(\sqrt 2  \k)^{{2\ov d} - {N\ov 2}}
 d^{- {d+1 \ov d}} \l^{1-\l} 
 L^p ({  q_p \ov  \sqrt N})^{N\ov 2}  E^{\l-1}  \ . }
Notice that the temperature does not depend on the energy if 
$\l=1$. This happens
if $N= 2/d$, i.e. if $d=1, \ N=2$ and also if  $d=2, \ N=1$.
The latter  case ($p= D-5$) corresponds to the {\it 5-brane } in $D=10$
or the fundamental string in $D=6$.
Thus, for the $D=10$ 5-brane we get a surprisingly simple expression
for the entropy
\eqn\fiff{
S_5^{(10)} = ({\sqrt 2 \k q_5})^{1/2}  E \ . }
Note that the explicit dependence on the volume, and all the factors of
$\pi$, have canceled out.
For the R-R charged 5-brane the string coupling  goes to zero 
at the  horizon.  Perhaps the fact that the entropy is linear in the
energy may be interpreted as the absence of 
thermodynamical equilibrium in this system
due to the vanishing coupling on the world volume.

For $\l \not= 1$ we may express the entropy in terms 
of the charge and the temperature,  
\eqn\kkk{ S_p \propto  L^{p}
 q_p^{N\ov 2(1-\l)}  T^{\l\ov 1-\l} \ . }
It is interesting to identify the cases where
the Bekenstein-Hawking entropy scales in the same
way as the entropy of an ideal gas of massless particles 
in $p$ dimensions (the gas of massless modes 
on the world volume), 
i.e. when $ S_p \sim T^p$. 
The condition $  {\l\ov 1-\l} = p$  is equivalent to
$p+1 = 2d/(Nd-2)$. 
It follows that $a=0$, i.e. that the dilaton is constant.
Then
\eqn\dila{ N= {2 \ov p+1}  + 
{2 \ov d}=  {2(D-2) \ov (p+1)(D-3-p)}\ , \ \ \ \ \  \l= {p \ov p+1} \ .  } 
We conclude that {\it for all the non-dilatonic $p$-branes 
the  entropy  has the natural massless ideal gas scaling}, 
$S_p \sim  L^p T^p$.
This is a reasonable conclusion: only if the coupling is 
well-behaved 
may we hope to reproduce black $p$-brane 
thermodynamics by 
a simple weakly interacting ensemble. 
The condition \dila\ is very restrictive in the supersymmetric cases: if 
 $N=1$  \dila\ is satisfied only for $p=2$ and $p=5$ in $D=11$
and for $p=3$ in $D=10$; if $N=2$ the only solution is $p=1$ in $D=6$.  

In what follows we shall specialize to non-dilatonic p-branes
\ght\ \dila, 
but keep $N$ arbitrary. Then  
\eqn\ee{
S_p= {4\pi  \omega_{d+1} \ov  2\k^2 } \left  ({4\pi \ov d}\right)^{p}
\left ({\sqrt 2  \k \ov \omega_{d+1} \sqrt N d } 
q_p\right )^{{ D-2\ov D-3-p}} L^p  T^p \ ,   }
or, equivalently, 
\eqn\eepe{
S_p= {4\pi  \omega_{d+1}  r_0^{d+1+p}\ov  
2\k^2 } ({4\pi \ov d})^{p} L^p  T^p \ .  }
The  charge thus enters in the power 
2 for the self-dual string in $D=6$ and 3-brane in $D=10$, 
in the power 3 for the 5-brane in $D=11$, and the power 3/2 for
the 2-brane in $D=11$. 
For the first two of these, the self-dual 
p-branes  ($d=p+1$, $N= 4/(p+1)$),
the expression for the entropy reduces to
\eqn\eeo{
S_p = {1\ov 2\omega_{p+2}  } ({4\pi \ov p+1})^{p+1} q_p^2 L^p  T^p \ .   }

%%%%%%%%%%%%%%%%%%%%%%%%%%%%%%%%%%%%%%%%%%%%%%%
\newsec{Charge quantization  conditions}
%%%%%%%%%%%%%%%%%%%%%%%%%%%%%%%%%%%%%%%%%%%%%%%%%%%
To carry out a detailed comparison
of the above expressions for the Bekenstein-Hawking
entropy 
with statistical mechanics, we need to decide how 
the $p$-brane charge is quantized. 
In the case  when both 
elementary   and solitonic 
p-branes  exist in the same theory 
the corresponding `electric' and `magnetic' charges
satisfy the  quantization condition ($\td d =p+1$) \teit\
\eqn\qua{
e_{\td d} g_d = 2\pi m  \ ,   }   
where $m$ is an integer. For the elementary charges this condition
is typically satisfied with $m=1$.

In the dyonic case, the quantization condition becomes
$e_{\d} g'_d - e'_{\d} g_d =2\pi m$. 
This does not a priori fix 
the value of $e_{p+1}=g_{p+1}=q_p$ in self-dual case. 
Here we need to envoke some extra information about 
how the quantum theory is actually defined.
For example, in the case of the dyonic string in $D=6$ 
one may  consider it being a special NS-NS solitonic string solution 
with quantization of charges being fixed  by  fundamental string 
world-sheet considerations (e.g. quantization of magnetic charge 
is fixed by  its relation to the WZW term in the string action \row\
or the level of the $SU(2)$ WZW model). The  resulting charge quantization
(which agrees with the one 
used in \hst) might differ from the one that applies to the genuine 
interacting  $D=6$ self-dual string.
In the 3-brane case the unit of charge is fixed 
by interpreting it as a $D$-brane configuration in type IIB superstring
\gubs.

Since for $D=2p+4$ the electric and the magnetic $p$-branes
are interchanged by the weak-strong coupling duality, we assume that
the elementary electric and magnetic charges are equal.
Using the quantization condition \qua\ 
with $m=1$, we find that the elementary unit of charge is
$\sqrt {2\pi}$. Thus, the possible values of charge are
$q_p=\sqrt{2\pi}  n $, where $n$ is the integer equal to the number
of coincident $p$-branes. We believe that this quantization rule is correct
for dyonic strings in $D=6$, and, in particular, for the special case of equal
charges (as mentioned above, 
this is justified by reference to fundamental string theory). 
For dyonic systems with equal charge, \eeo\ becomes 
\eqn\self{
S_p= {\pi \ov \omega_{p+2}  } \left ({4\pi \ov p+1}
\right )^{p+1} n^2 L^p  T^p \ .  }
For example, for the $D=4$ $0$-brane and the $D=6$  $1$-brane we 
find, respectively, 
\eqn\see{
S_0 = {\pi}   n^2  \ , \ \ \ \  \ \ S_1^{\rm dyonic} = 2\pi  n^2 L  T  \ . }
This result for the string agrees with the statistical count
of entropy based on D-branes \hst.

At the same time, the $D$-brane interpretation of 3-branes implies \gubs\
a quantization condition which is  different by a factor of 2
(this `compensates' for $1/\sqrt 2$ factor in the expression for the charge \mess\ in the dyonic case)
\eqn\uuuy{ q^2_p   = \pi  n \ .  } 
Then 
\eqn\seei{
S_p= {\pi \ov 2 \omega_{p+2}  } ({4\pi \ov p+1})^{p+1} n^2 L^p  T^p \ .  }
We believe that this result is correct both for the self-dual 3-brane
and for the `truly self-dual' string, 
\eqn\sdentropy{
 S_1^{\rm self-dual} = \pi  n^2 L  T \ , \ \ \ \ \ \ \ 
S_3= {\pi^2 \ov 2} n^2 L^3  T^3 \ .  }
Thus, there may be a subtle difference
between the equal charge case of the dyonic string, considered in
\hst, and the `truly self-dual' string. Clearly, this deserves 
further investigation.

Let us also note that, in addition to the obvious similarities, there are
also obvious differences between dyonic
strings in $D=6$   and 3-branes in $D=10$.
For the former, the tension is given in terms of the 
integral electric and magnetic charges, $P$ and $Q$, by
$( e^{-\p_0} P + e^{\p_0} Q)/2\pi \a'^2$. For the equal-charge dyonic 
string,
$P=Q$ 
and $e^{\p_0} =1$,  which implies that there is no freedom to adjust the
string coupling. This is also consistent with the absence of the dilaton
field in the six-dimensional self-dual gravity.

For the self-dual 3-brane there is no `dyonic' generalization with
$P\neq Q$ because it is stabilized by the self-dual 5-form.
Another apparent difference is that the dilaton may assume any constant
value. As we remarked earlier, it is not obvious, however,
that there is a limit
where the world volume theory of multiple 3-branes is truly weakly coupled.
The charge quantization subtleties that we have discussed suggest
that, even when classical solutions are identical, there may be a difference
between `truly self-dual' objects and equal-charge dyonic objects.

Let us now turn to the case of  our main interest --  $p$-branes 
in $D=11$ supergravity or `M-theory'.  We shall assume  
that  there exist both 
the fundamental 2-branes and the solitonic 5-branes
with their unit charges related by 
\eqn\unitcharge{q_2 q_5= 2\pi \ .} 
The unit 2-brane charge is fixed in terms of the 
tension  $T_2$ in the elementary 
2-brane action, which plays the role of  a source 
in supergravity equations of motion, 
\eqn\yry{
q_2  = \sqrt 2 \k T_2 \ . }
This normalization is 
consistent with the fundamental string one. Assuming that
 2-brane is wrapped around 2-torus with periods $L$
we may use double dimensional reduction  \duf\
to  relate the membrane  solution
to the fundamental string solution \duh.
It is easy to see that  the string 
and membrane tensions, $T_1 = 1/2\pi \a'$ and $T_2$,
are related by 
\eqn\doubled{T_2= T_1/L\ , 
\qquad\qquad T_2 \k^2 = T_1 \k^2_{10} \ ,}
where 
the 10-dimensional gravitational constant is expressed in terms of  
the 11-dimensional one by $\k^2_{10}= \k^2/L$.
If the string is wrapped $n$ times around the 9-th dimension
the coefficient in the corresponding harmonic function
is $
r_-^6 = {2 \k^2_{10}\ov  6 \om_7} n T_1  = {2 \k^2\ov  6 \om_7} n T_2$ .

In order 
to determine $T_2$ in terms of $\k$ we need one more input.
We will use the fact that the double dimensional reduction of
the M-theory 5-brane gives the Dirichlet 4-brane of the type IIA theory,
\eqn\doubledim{T_5 \k^2 = T_4 \k^2_{10} \ .}
The unit 5-brane charge is related to $T_5$ by
$q_5=\sqrt 2 \k T_5$.
Using $\k_{10}=g \alpha'^2$, eqs. \unitcharge\--\doubledim, and the
known D-brane tension \polch, $\k_{10} T_4= (2\sqrt{\pi\alpha'})\inv $, we 
solve for $\k$ in terms of the ten-dimensional quantities,
\eqn\plan{\k^2= {g^3 \alpha'^{9/2}\over 4 \pi^{5/2} }\ .}
Our argument further determines the charges and tensions of the M-branes.
In particular,\foot{In ref. \mina\ it was argued on the basis
of membrane path integrals that $\k^2 T^3_2 = \pi^2/m_0$.
The quantity $m_0$ was left ambiguous, however. Our line of reasoning based on 
relation to $D=10$ string theory 
fixes the value of $m_0$ to be $1/2$. It would be
interesting to find an intrinsic
 M-theory argument which explains 
why  $\k^2 T^3_2 = 2\pi^2$ is the correct quatization condition.}
\eqn\tetwo{ T_2^3= {2\pi^2\over \k^2}\ .}
Thus, the  2-brane charge  takes on the following values, 
\eqn\yri{
q_2  =  n \sqrt 2  (2\k \pi^2)^{1/3}  \ .  }
The quantization condition \qua\  fixes  the possible values of the 
5-brane charge  to be  
\eqn\yik{
q_5  = {2\pi n \ov \sqrt 2 \k T_2} =  n \sqrt 2({\pi\ov 2\k})^{1/3}  \ . }
Substituting these expressions into the entropy \ee, we find 
for $n$ coincident 2-branes in $D=11$ ($p=2, \ d=6, \ \om_7= \pi^4/3$),
 \eqn\eet{
S_2^{(11)} = 2^{7/2} {3}^{-3}   \pi^2  n^{3/2}  L^2  T^2 \ .   }
For $n$ coincident 5-branes in $D=11$ ($p=5, \ d=3, 
\ \om_4= 8\pi^2/3$), 
\eqn\eef{
S_5^{(11)}  = 2^{7} {3}^{-6} \pi^3  n^3 L^5  T^5 \ .  }

%%%%%%%%%%%%%%%%%%%%%%%%%%%%%%%%%%%%%%%%%%%%%%%%%%%%%%%%%
\newsec{Statistical entropy of massless modes on $p$-branes
and comparison to the Bekenstein-Hawking entropy}
%%%%%%%%%%%%%%%%%%%%%%%%%%%%%%%%%%%%%%%%%%%%%%
Let us take a broader point of view than the one suggested by
the D-brane picture
(cf. \refs{\das,\CM,\hst,\gubs}).
We simply ask whether
it is possible to understand the entropy of  near-extremal 
non-dilatonic black p-branes  
in terms of the degeneracy of some weakly interacting massless 
$p$-brane excitation modes.
The corresponding free energy and entropy 
will be that of an ideal gas of massless particles in $p$
spatial dimensions. The number of boson and fermion fields is, of course,
determined by the detailed properties  of  
a particular $p$-brane.
 
%%%%%%%%%%%%%%%%%%%%%%%%%%%%%%%
\subsec{Massless ideal gas}
%%%%%%%%%%%%%%%%%%%%%%%%%%%%%%%%%%%
The free energy of massless Bose gas in $p> 0 $ spatial dimensions 
identified with periods $L$ is, in the thermodynamic limit of large $L$, 
\eqn\gas{
-F_p/T =  \ln Z_p =  L^p \int {d^p k \ov (2\pi)^p}  \ln ( 1 - e^{- |k|/T}) = c_p  L^p T^p \ , } $$
 c_p = (2\pi)^{-p}  \omega_{p-1} (p-1)! \ \zeta (p+1)  \ .  $$
The corresponding entropy and energy 
are given by 
\eqn\ene
{ S_p = - {\del  F_p \ov \del T} = (p+1) c_p  L^p T^{p} \ , 
\ \ \  \ E =  T^2 {\del  \ln Z_p  \ov \del T}  = p c_p  L^p T^{p+1} \ . }
The statistical treatment of a massless fermion field is analogous, 
with the conclusion that
one real fermion contributes the entropy equal to
$(1- 2^{-p})$ times that of a massless boson.
If the extremal $p$-brane preserves some supersymmetry, then
the number of massless bosons should equal the number of
massless fermions. Denoting this number by $N_p$, we find that
the total statistical entropy is
\eqn\enn{
S_p = b_p L^p T^{p} \ , \ \ \ \ 
b_p =  N_p[1+(1 - 2^{-p})]  (p+1)  c_p  \ . }
For the particular cases of a string, a 3-brane, and a 5-brane we find
\eqn\oop{ b_1 = { \pi\ov 2}  N_1 \ , 
\ \ \   \ 
b_3 = {\pi^2 \ov 12} N_3 \ , \ \ \ \ 
\ \ 
b_5= {\pi^3 \ov 40} N_5  \ . }
For the 2-brane, on the other hand,
we find $b_2= {7 \ov 8 \pi } \zeta(3) N_2$.
In general, the coefficients $b_p$ contain $\zeta(p+1)$, which 
for even $p$ are not expressible
in terms of $\pi$.
Thus, for even $p$ there cannot be a detailed
agreement between the Bekenstein-Hawking and the statistical entropy.
This may be related to the fact that the even $p$ solutions are
singular, while the non-dilatonic odd $p$ solutions are not \ght.
It is nevertheless intriguing that for the 2-brane in 11 dimensions
the scaling of the entropy with respect to the temperature
does agree with the massless statistical mechanics.

For the non-dilatonic $p$-branes with $p$ odd, we hope to find a more
detailed agreement. Indeed, comparing \oop\ with
the Bekenstein-Hawking entropies calculated in section 3, we find that
the powers of $\pi$ match perfectly. For the equal-charge (self-dual)  case of 
the dyonic  string in $D=6$,
we need $N_1= 4n^2$ to achieve perfect agreement. This is indeed
the number predicted by the D-brane counting of \hst. The factor of 4
may be thought of as the number of transverse 
modes of a string in 6 dimensions. The advantage
of our approach to charge quantization
and the normalization
of the entropy is that, although somewhat heuristic, it could be
carried out without any knowledge of the D-branes.

To achieve perfect agreement for the self-dual 3-brane, we need
$N_3= 6n^2$. As suggested in \gubs, it is tempting to think of the
factor 6 as the number of transverse modes of a 3-brane in 10 dimensions.
It is remarkable
that the $n^2$ growth is in agreement with the enhanced $U(n)$ symmetry
of $n$ coincident 3-branes \gubs. 
Naively, the number of massless field modes
is $8 n^2$, but massless monopoles and dyons are, 
in a sense, also present. Perhaps
$8 n^2$ is reduced to $6n^2$ by a confinement mechanism.

%%%%%%%%%%%%%%%%%%%%%%%%%%%%%%%%%%%%%%%%%%%%%%%%%%
\subsec{The 5-brane in D=11}
%%%%%%%%%%%%%%%%%%%%%%%%%%%%%%%%%%%%%%%%
For $n$ coincident $D=11$ 5-branes we need
\eqn\need{N_5=2^{10} 3^{-6} 5 n^3}
to achieve perfect agreement between the statistical and the
Bekenstein-Hawking entropies. While we certainly do not have enough
knowledge about the world volume theory to derive this number of
massless degrees of freedom, we have several comments.

The most notable feature, which does not depend on the details of how
the charge is quantized, is the $n^3$ growth of $N_5$. This is consistent
with the idea of enhanced symmetry for coincident 5-branes, but grows
faster than the dimension of any Lie group. Perhaps this faster growth
than the one found for D-brane systems is related to the appearance of
tensionless strings \refs{\ast,\wit}. 
Based on our entropy considerations, we speculate
that the enhanced symmetry in M-theory is a new kind of phenomenon
that needs to be understood better.

Unlike in the 1-brane and 3-brane cases, the number of massless degrees
of freedom is in general fractional, but becomes integer when
$n$ is a multiple of $9$. It is conceivable that the fractional effective
number of massless fields is a regularized version of the infinite number
of massless fields which appear when 
the closed string in $5+1$ dimensions becomes tensionless.
Furthermore, the world volume theory is not expected to be weakly coupled,
which further complicates matters.

%We should also remark that, although our method to determine the unit of
%5-brane charge seems sound, there may be some subtleties.

If we look for a pattern in the entropy results, it seems that
the correct factor might simply be the number of transverse modes.
What could be the possible justification for ignoring the longitudinal
modes, which in the D-brane description are the world volume gauge fields?
While we do not know the answer, we note that the gauge modes are connected
with the antisymmetric tensor fields in the bulk. Indeed, the
world volume gauge fields may be thought of as gauge transformation parameters
for $B_{\mu\nu}$. For this reason it may be redundant to count these modes.
Of course, we hope that there is a good dynamical justification for
excluding them.
For the 5-brane in $D=11$ there are 5 transverse modes, and by analogy
we might have expected that $N_5= 5 n^3$. The extra factor we are finding,
$2^{10} 3^{-6}$, is numerically close to 1 and may one day find an
explanation.

While there are many puzzles remaining, we feel that the close
correspondence between the Bekenstein-Hawking and the statistical
entropies for a 5-brane in 11 dimensions are unlikely to be
a mere coincidence. Both the scaling with respect to $T$ and the
power of $\pi$ match, suggesting the presence of massless fields
on the world volume. We hope that the detailed counting teaches
us something about the effective action for the world volume theory.

%%%%%%%%%%%%%%%%%%%%%%%%%%%%%%%%%%%
\newsec{Concluding remarks }
%%%%%%%%%%%%%%%%%%%%%%%%%%%%%%%%%%%%%%%%

In this paper we discussed the entropies of several non-dilatonic
$p$-brane systems: the dyonic string in $D=6$, the self-dual
$3$-brane in $D=10$, and the $2$- and $5$-branes in $D=11$.
For all these cases, and especially for the odd-dimensional branes,
there is a close correspondence between the Bekenstein-Hawking entropy
and the statistical entropy of massless fields on the world volume.

We cannot resist taking this one step further and offering some
vague speculations about the 5-brane in $D=12$. Although there is no
known supergravitySubst for $D>11$, there has been some recent evidence in 
favor of the existence of a
12-dimensional theory \refs{\hull,\vafa,\kut}. 
In particular, the important role played by the 3-brane of type IIB
theory suggests a possibility of its 12-dimensional reformulation \aat.
Since the elementary string in $D=10$, the elementary 2-brane in $D=11$,
and the elementary 3-brane in $D=12$ are all dual to the 5-brane,
this object seems to have a special role.
If we assume that there is a non-dilatonic 5-brane solution in $D=12$,
and apply a very heuristic charge quantization procedure based on comparing
it with the 3-brane, we find that
the Bekenstein-Hawking entropy is quantized in units of
\eqn\kep{
S^{(12)}_5=  { 6 \ov 5 }     \pi^3   L^p  T^p \ .    }
Amusingly, this corresponds
to $N_5= 6 \times 8$ degrees of freedom on the world volume.
One may speculate that $6$ is due to the number of transverse modes
of the 5-brane in 12 dimensions, while $8$ is related to the number
of world sheet degrees of freedom of the self-dual string.
In any case, unlike for $D=11$, we seem to find an integer number
of massless fields on the world volume. While at the moment we can
offer little more than this kind of numerology, we believe that further work
will shed more light on the dynamics of $p$-branes in the
hypothetical 12-dimensional theory.

\newsec{Acknowledgements}
We thank C. Callan, R. Dijkgraaf, E. Kiritsis and E. Verlinde for interesting
discussions.
I.R.K. was supported in part by DOE grant DE-FG02-91ER40671, the NSF
Presidential Young Investigator Award PHY-9157482, and the James S.{}
McDonnell Foundation grant No.{} 91-48.
A.A.T. would like to acknowledge  the   support of PPARC,
ECC grant SC1$^*$-CT92-0789 and NATO grant CRG 940870.

\vfill\eject
\listrefs
\end